\documentclass[twocolumn, amssymb, prl, epsfig, superscriptaddress]{revtex4}
\usepackage[utf8]{inputenc}
\usepackage{color}
\usepackage[normalem]{ulem}
 
\usepackage{amssymb}
\usepackage{amsmath}

\usepackage{graphicx}
\usepackage{epsfig}
\usepackage[pdftex, bookmarks,colorlinks=true,urlcolor=blue]{hyperref}

\begin{document}

\title{Collective transport of charges in charge density wave systems based on traveling soliton lattices}

\author{A. Rojo-Bravo}

\author{V.L.R. Jacques}

\author{D. Le Bolloc'h}

\affiliation{Laboratoire de Physique des Solides, Universit\'e Paris-Sud, CNRS, UMR 8502, F-91405 Orsay, France}

\begin{abstract}
Solitons are peculiar excitations that appear  in a wide range of nonlinear systems such as in fluids or optics. We show here for the first time that  the collective transport of charges observed in charge density wave (CDW) systems can be explained by using a similar theory based on a traveling soliton lattice. Coherent x-ray  diffraction experiment performed in the sliding state of a CDW material reveals peculiar diffraction patterns in good agreement with this assumption. Therefore, the collective transport of charges  in CDW systems may be due to a nonlinear interaction leading to a self-localized excitation, carrying charges without deformation through the sample, on top of the CDW ground state. This single theory  explains why charges remain spatially correlated over very long distances and reconciles the main features of sliding CDW systems,  either  observed by transport measurements or diffraction.
\end{abstract}
\maketitle
A soliton can take the form of a  localized solitary wave which propagates in a medium while keeping a constant shape through nonlinear interactions. Once created, this  localized wave, with particle-like properties, propagates without dispersion and  with a remarkably large lifetime.  Solitons are present in many systems such as fluids\cite{dauxois_physics_2006}, or optical fibers\cite{RevModPhys.68.423} but also in  more unexpected fields  like traffic jams\cite{nagatani_physics_2002} or blood pressure\cite{yomosa_solitary_1987-2}. Their involvement in electronic crystals is invoked in only few systems, like Josephson junctions\cite{Lomdahl} and conducting polymers \cite{RevModPhys.60.781}, manifesting itself in different forms. 


CDW systems are another type of electronic crystals made of spatially correlated electrons.  CDW can coexist with spin density wave like in chromium\cite{fawcett_spin-density-wave_1988, PhysRevB.89.245127} or compete with superconductivity like in cuprates\cite{Ghiringhelli:2012bw}.
Although the static CDW state is now well understood, the dynamical one is still debated. Indeed, the most spectacular property of a CDW system  is its ability to carry correlated charges when submitting the sample to an external electric field.  Above a threshold field E$_{th}$, a non-ohmic resistivity is observed, including voltage oscillations with a fundamental frequency $f_0$ proportional to the applied field, as well as several harmonics  in the frequency spectrum\cite{Fleming:1979cu}. Up to 23 harmonics have been observed in NbSe$_3$\cite{thorne_charge-density-wave_1987}. This evidence of  collective transport  through CDW systems has received considerable interest for more than 35 years. However, the understanding of the type of charge carriers and their propagation mode remains incomplete. 

Many theoretical approaches have been proposed  to describe this phenomenon. The simplest one is based on translation invariance of the incommensurate CDW\cite{Frohlich:1954fz}: the whole sinusoidal  density of  condensed charges {\it slides} over the atomic lattice with a constant velocity. Although appealing, this explanation remains probably approximate since the CDW is described as an almost sinusoidal modulation   from diffraction experiments\cite{note1} while transport measurements reveal a strong anharmonic signal\cite{thorne_charge-density-wave_1987}.
A more realistic description of CDW dynamics assumes a slowly varying phase $\phi$(x) of the CDW interacting weakly with impurities\cite{Fukuyama:1978br}. On the contrary, theories considering strong electron-phonon coupling neglect the role of impurities and treat CDW dynamics as only  due to phonons\cite{aubry_low-dimensional_1989}. The most accepted theory, developed by Ong and Maki\cite{Ong:85om} and Gor'kov \cite{Gorkov:1983tb, Gorkov:1984vj, Batistic:1984by}, deals with the CDW-metal junction at  electrical contacts.  The conversion of normal electrons from the metallic electrode into  condensed charges in the CDW is made possible by climbing CDW dislocations at the interface.
These so-called {\it phase slippage} and current conversion phenomena are in agreement with local resistivity measurements  close to contacts\cite{maher_charge-density-wave_1995}.

This conversion phenomenon is also accompanied by a static and an elastic deformation of the CDW. Without external field, the CDW, $\rho=\rho_0\cos(2k_F x+\phi(x,t))$, is homogeneous along the sample and pinned at both ends at the metal/CDW junctions. In the sliding state however, the CDW is compressed at one electrode and stretched at the other as revealed by diffraction experiments\cite{dicarlo_field-induced_1993,Requardt:1998jz}. The corresponding time-averaged phase follows a quadratic behavior:$\left< \frac{\partial^2 \phi}{\partial x^2} \right>_t =cst$\cite{Feinberg:1988hz}.

\begin{figure}
\includegraphics[width=.45\textwidth]{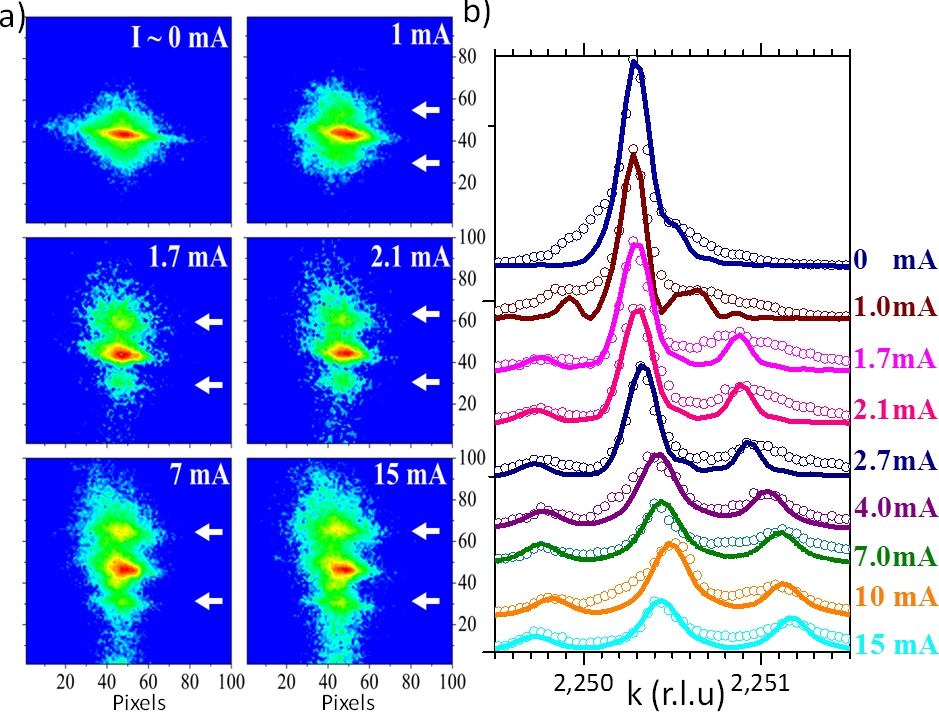}
\caption{Coherent diffraction patterns of the 2k$_F$ satellite reflection associated to the CDW versus applied currents. (a) The $ \text{Q} \, = (1, q_{2kF}, 0.5)$ with $q_{2kF}\approx 0.748 \, \text{b*}$ satellite reflection in the blue bronze K$_{0.3}$MoO$_3$ for several external currents (T $= 70$ K). The 2D patterns corresponds to a sum over several $\theta$ angles through the maximum of intensity. The b* chain axis  is vertical and the additional satellites are indicated by white arrows. The scale is in pixels  (logarithmic color scale from blue for lowest intensity to red for highest intensity). (b) Corresponding profiles along b* after summation along the perpendicular axis (opened circles) and fits using the soliton lattice model (continuous lines). A convolution with a Lorentzian shape has been used to take into account the finite Q-resolution as well as the projection along the chain axis b*, the integration over several $\theta$ angles and  temporal fluctuations during data acquisition.}
\label{fig:peak}
\end{figure}

The impressive number of studies focusing on sliding CDW deserves, however, a few comments. While many studies have been devoted to the conversion process close to electrodes, the  charge carriers  propagation through macroscopic samples remains a subject that has been little studied to date. A pure quantum tunneling through the sample was mentioned\cite{Bardeen:1979hv} and phase slippage mixed with quantum tunneling have also been considered at low temperature\cite{MAKI:1995wu}. However, current oscillations are clearly observed in very long samples, up to several centimeters for NbSe$_3$ and a pure quantum tunneling over such large distances is probably unlikely.
Note  also that, in the phase slippage theories\cite{Ong:85om}, impurities play a minor role, hidden in the tunneling coefficient. The authors justify this absence  by the increase of  CDW correlation lengths $\xi_l$ in the sliding regime. Nevertheless, recent diffraction experiments  show that $\xi_l$ is always shorter  in the sliding state than  in the pristine one\cite{Pinsolle:2012dy}, suggesting on the contrary that defects may still  play a  role in the sliding state. 

Several  ascertainments can also be done with respect to diffraction experiments. First, CDW systems can stabilize CDW dislocations, i.e. abrupt phase shifts of the CDW modulation, first observed by coherent x-ray diffraction\cite{le_bolloch_charge_2005}. Second, those CDW dislocations are mobile in the sliding regime, as proved by the disappearance of speckles\cite{Pinsolle:2012dy}, suggesting that moving phase shifts can play a role in charge transport.

The starting point of this study is the reinterpretation of experimental data we have obtained in the archetype K$_{0.3}$MoO$_3$ blue bronze  system under applied  currents\cite{le_bolloch_observation_2008}. The most spectacular point of this experiment is the appearance of two secondary satellites located on both sides of the 2k$_F$ satellite reflections associated to the CDW (see Fig.1). The corresponding spatial frequencies leading to micron-size distances have been observed thanks to  coherence properties of the x-ray beam\cite{jacques_bulk_2011}. In \cite{Jacques:2012ex}, we used a phenomenological static model to account for this observation. We develop  here a fundamentally different model, based on the existence of a dynamics 2$\pi$-solitons lattice,
able to account for not only this coherent diffraction experiment but also for the all features of a sliding CDW discussed in the introduction,  either observed  by transport measurements or diffraction.


\begin{figure}
\includegraphics[width=.45\textwidth]{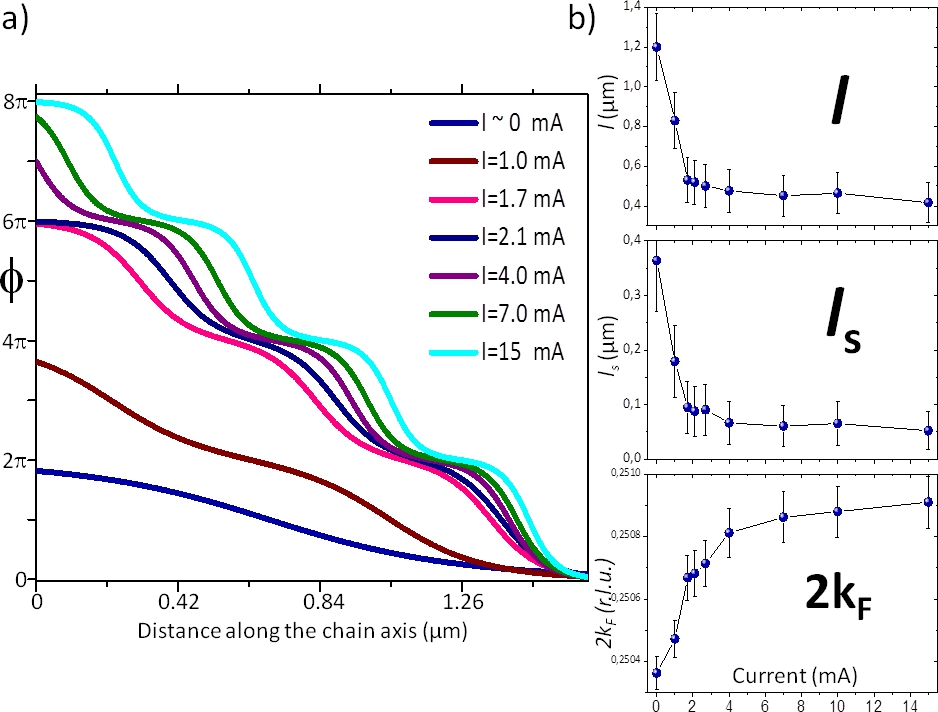}
\caption{(a) Profile of the CDW phase $\phi(x)$ obtained from the fits of  Fig.1b versus the distance along the chain axis  for different external currents. (b) Distance $l$ between solitons,  soliton size $l_S$ and  2k$_F$  versus currents.}
\label{fig:soliton_lattice}
\end{figure}
An interaction that couples impurity potential and the phase $\phi$ of the CDW  is considered as in \cite{fukuyama_electronic_1985}. The corresponding phase-dependent Hamiltonian leads to the following equation of motion:
\begin{equation}
\label{eq:theosol004}
\frac{\partial^2 \phi}{\partial t^2} - c_\phi^2 \frac{\partial^2 \phi}{\partial x^2} + \eta \frac{\partial \phi}{\partial t}+\omega_0^2 \sin \left( \phi \right) = F 
\end{equation}
where $F=\frac{2 c_\phi^2 \, e \, E}{\hbar \, v_F}$ is proportional to the applied force,   $c_\phi = \sqrt{m / m^*} \, v_F$ is the phason's velocity and  $\omega_0$ the pinning frequency. We also add an effective damping term $ \eta \, \frac{\partial \phi}{\partial t}$ to mainly take  into account  the coupling between CDW and phonons. Contrary to \cite{fukuyama_electronic_1985}, the $\sin \left( \phi \right)$ term is not linearized allowing abrupt phase variations. The usual non-perturbed sine-Gordon equation (for which F=$\eta$=0) is known to admit soliton solutions. However,  soliton excitations are quite robust and survive the inclusion of a reasonable external force and dissipation keeping their topological properties although the soliton shape is slightly modified\cite{Fogel:1977bw}. 

Let's now solve  Eq.1 considering that the  phase $\phi(x,t)$ contains two terms: a slowly varying phase $\phi_0(x)$ and a dynamical part  $\phi_1(x,t)$ where $\phi_1$ varies much more rapidly than the static one ($\langle|d^2 \phi_1 / dx^2|\rangle_t \gg \langle|d^2 \phi_0 / dx^2|\rangle_t$). The static part $\phi_0(x)$ can be calculated by  averaging Eq.1 in time :
\begin{equation}
\label{eq:theosol008}
\left<  \frac{\partial^2 \phi_0(x)}{\partial x^2} \right>_t =(\frac{\eta \, \pi}{e} j- F)/c_\phi^2,
\end{equation}
where the excess of current  in the sliding mode $j =\frac{e}{\pi} \frac{\partial \phi}{\partial t}=\frac{e}{\pi}v_S$ is constant  far from  electrodes as observed by several transport measurements\cite{MAHER:1993ib, GILL:1993bq, Adelman:1995ca, ITKIS:1997jf, Lemay:1998wx}.
This leads to a quadratic variation of the  phase $\phi_0(x)$  in perfect agreement with diffraction experiments\cite{dicarlo_field-induced_1993,Requardt:1998jz}.
The complete Eq.1 now reads:
\begin{equation}
\label{eq:theosol018}
\frac{\partial^2 \phi_1}{\partial t^2} - c_\phi^2 \frac{\partial^2 \phi_1}{\partial x^2} + \omega_0^2 \sin \left( \phi_1 \right) = F + c_\phi^2 \frac{\partial^2 \phi_0}{\partial x^2}- \eta \frac{\partial \phi_1}{\partial t}.
\end{equation}
The dynamical part $\phi_1(x,t)$ obeys the sine-Gordon equation and is submitted to an effective force including the friction.
Considering the periodic nucleation of CDW dislocations at the electrode\cite{MAKI:1995wu}, we obtain a train of solitons plus a negligible quantity $\delta$\cite{Fogel:1977bw}:
\begin{equation}
\label{eq:theosol019}
\phi_1(x,t) = \delta+ \sum_{n=-\infty}^\infty 4 \arctan \left( \exp \left( \frac{x - v_S t - l n}{l_S \, \gamma (v)} \right) \right),
\end{equation}
where $l$ is the distance between successive solitons and $l_S= c_\phi / \omega_0$ their extension. Overlapping effects between solitons are neglected ($l / l_S > 2$).

\begin{figure}
\includegraphics[width=.45\textwidth]{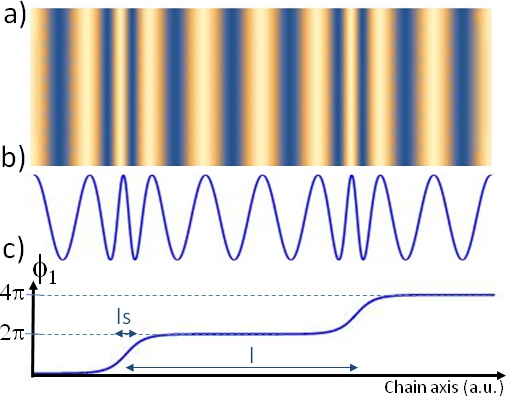}
\caption{Sketch of the static  soliton lattice in real space. (a) CDW in the presence of a soliton lattice with  $l/l_S=20$. The lighter wave fronts correspond to an excess of electrons with (b) the corresponding electronic density profile. (c) The corresponding phase $\phi_1$. The movie of the moving soliton lattice can be seen in the Supplementary Information.}
\label{fig:simulation}
\end{figure}

The soliton lattice model presented here leads to a singular diffraction patterns in good agreement with the experiment, especially for larger currents when the soliton lattice is well formed (see Fig.1). Two additional satellites reflections appear on both sides of the main 2k$_F$ peak located  at $\delta q =\pm 2\pi / l$. Since Eq. \ref{eq:theosol019} is not an even function, the two satellites at $\delta q$ do not have the same intensity in agreement with the experiment. The soliton extension $l_S$ mainly affects the intensity ratio between the central peak and the two satellites. Note that the central peak is not necessarily  located at 2k$_F$ but  may be shifted with respect to $l/l_S$\cite{note3}. Note also that the soliton model used here is global in the sense that one soliton can not  be considered individually without considering the complete soliton lattice. The three main fitting parameters ($l$, $l_S$ and  2k$_F$) are extremely sensitive to each other and the solution space is particularly narrow. This is a good guaranty of the unicity of the solution.

The traveling soliton lattice correctly accounts for the diffraction patterns (see Fig. 1a). The corresponding phase $\phi(x)$ and the behavior of $l$, $l_S$ and 2k$_F$ versus external current are shown in Fig.2. The distance between solitons $l$ reaches the micron size for small currents,  decreases for $I < 2 I_S$ and  reaches a stationary value above  2I$_S$ where I$_S$ is the threshold current.
 The soliton extension $l_S$ follows a similar behavior versus current. The distance $l$ between solitons is always  greater than the soliton extension $l_S$ ($l / l_S >3$) which  justifies the assumption of non-interacting solitons.

In this non-linear framework, the existence of a finite threshold field $E_\text{th} $ (for which $\frac{\partial^2 \phi_1}{\partial t^2}=\frac{\partial \phi_1}{\partial t}=0$ in Eq.1) is obvious, in agreement with the experiment. For $E / E_\text{th} \gg 1$, the soliton reaches a stationary sliding velocity  \cite{Fogel:1977bw},
\begin{equation}
\label{eq:theosol015}
v_S = \frac{\pi \text{ e } \tau^* v_F k_F}{2 \omega_0 m^*} \sqrt{\frac{m}{m^*}} E,
\end{equation}
proportional to the applied field $E$, in agreement with the experiment ($v_F$ is the Fermi velocity and $m^*$ the effective mass). In this framework, a quantitative sliding velocity $v_s$ can be given. Since the observed $l$ saturates for large enough fields (see Fig. 2b) and the fundamental frequency ranged from $f_0=$1 Mhz to 100 Mhz\cite{Monceau:2012dg}, $v_S=f_0 l$ ranges from $v_S$=0.5 m/s to 50 m/s. 

Remarkably, the most impressive feature of a sliding CDW is the presence of many harmonics in the frequency spectrum as observed by transport measurements\cite{thorne_charge-density-wave_1987}. This signature is easily explained by the non-harmonic shape of the soliton when $l_S\ll l$ (see Fig. 2). Note that, in the opposite case, when $l_S \sim l$, the electronic density is close to a harmonic modulation and the sliding process is more similar to a global CDW translation as in the Fr\"{o}hlich approach\cite{Frohlich:1954fz}. 

In this semiclassical description,  charges are carried by phase shifts of the CDW modulation (see Fig.3)  which can travel through  macroscopic samples without deformation on top of the CDW ground state. The traveling 2$\pi$ soliton lattice which is superimposed to a slowly-varying static phase reconciles seemingly contradictory results:  on one hand, it explains results obtained by diffraction including  the macroscopic elastic deformation along the sample and the singular coherent diffraction patterns in Fig. 1. On the other hand, this single theory also explains the main features observed in transport measurements such as the existence of a threshold field, the presence  of correlated charges despite the large distances involved in this phenomenon and finally the presence of several harmonics in the frequency spectrum. The model presented here highlights a new type of charge transport based on traveling CDWs phase shifts and opens new perspectives in controlling correlated charges over macroscopic distances.

\bibliographystyle{unsrt}
\bibliography{./solitonbib13}

\begin{thebibliography}{10}
\bibitem{note1}
{Diffraction experiments display satellite reflections located close to 2k$_F$.
  Only very weak second harmonic peaks at 4k$_F$ can be observed in most CDW
  systems like NbSe$_3$, K$_{0.3}$MoO$_3$} or TbTe$_3$ showing that the CDW
  modulation is mostly sinusoidal.
\end{thebibliography}

\end{document}